# Spectral Visualization of Excitonic Pair Breaking at Individual Impurities in Ta$_2$Pd$_3$Te$_5$

Lianzhi Yang[1,9], Deguang Wu[2,3,9], Hanbo Zhang[1], Yao Zhang[1], Xiutong Deng[4,5], Chao Zhang[1], Tianyou Zhai[6], Wenhao Zhang[1,7], Youguo Shi[4,5], Rui Wang[2,3,8,*], Chaofei Liu[1,†] & Ying-Shuang Fu[1,7,§]

[1]School of Physics and Wuhan National High Magnetic Field Center, Huazhong University of Science and Technology, Wuhan 430074, China
[2]National Laboratory of Solid State Microstructures, Nanjing University, Nanjing 210093, China
[3]Collaborative Innovation Center for Advanced Microstructures, Nanjing University, Nanjing 210093, China
[4]Beijing National Laboratory for Condensed Matter Physics, Institute of Physics, Chinese Academy of Sciences, Beijing 100190, China
[5]Center of Materials Science and Optoelectronics Engineering, University of Chinese Academy of Sciences, Beijing 100049, China
[6]State Key Laboratory of Materials Processing and Die & Mould Technology, and School of Materials Science and Engineering, Huazhong University of Science and Technology, Wuhan, Hubei 430074, China
[7]Wuhan Institute of Quantum Technology, Wuhan 430206, China
[8]Jiangsu Physical Science Research Center, Nanjing University, Nanjing 210093, China
[9]These authors contributed equally: Lianzhi Yang, Deguang Wu

*rwang89@nju.edu.cn
†cliu@hust.edu.cn
§yfu@hust.edu.cn

Excitonic insulators host the condensates of bound electron–hole pairs, offering a platform for studying correlated bosonic quantum states. Yet, how macroscopic coherence emerges from locally collapsed pairing remains elusive. Here, using scanning tunnelling spectroscopy, we report the impurity-induced pair breaking in an excitonic insulator Ta$_2$Pd$_3$Te$_5$. Individual Te vacancies are found to generate a pair of spectral peaks within the excitonic gap. Their energies depend sensitively on the defect configurations and are continuously tunable by tip electric field, indicating controllable impurity scatterings. Spectral mapping shows spatially anisotropic and electronically coupled electron–hole components of the subgap states. These observations, together with mean-field modelling, suggest an excitonic pair-breaking origin. In the strongly electron–hole imbalanced region, a secondary pair-breaking effect, manifesting as an additional pair of subgap states with distinctly lower energies, can emerge, presenting the interplay of pairing breakings with different excitonic order parameters. Our findings demonstrate the spectroscopic 'fingerprint' of local excitonic depairing at the atomic level, offering a crucial clue to the critical behavior across excitonic condensation.

The dilute Bose gases that are condensed into the statistically degenerate ground state can lead to the fascinating Bose–Einstein condensates (BECs)[1]. Owing to the fragility to thermal fluctuations, such BEC, initially proposed in cold atoms, occurs at extremely low temperatures of ~μK[2], and is thus challenging to realize. One alternative solid-state platform for achieving BEC has been the excitonic insulator (EI)[3,4], which can enable the quantum condensation of electron–hole (e–h) pairs, or 'excitons' at significantly higher temperature (~10$^2$ K)[5,6]. Over decades after the original theories, the pursuit of EIs has expanded the studies to diverse systems, such as e–h bilayers[7-10], Ta$_2$NiSe$_5$[5,11], and 1$T$-TiSe$_2$[12,13]. Experimentally, signatures of the excitonic condensation are typically inferred from the transport signals arising from the collective binding of electrons and holes[14-17], or, spectroscopically, the hybridization gap opened due to excitonic band renormalizations[12,18]. Although extensive excitonic signatures have been reported, it still remains unknown how the globally coherent condensate is developed from unpaired carriers.

Impurity-induced pair-breaking state provides a unique opportunity for the dilemma. For example, in superconductors, the bosonic condensates akin to EIs, impurities have been shown generating pair-breaking states that reveal the preformation of Cooper pairs and their evolution to global coherence[19]. Theoretically, a similar pair-breaking phenomenon was proposed for EIs dating back to 1967, which predicted the localized impurity states within excitonic gap[20]. Nevertheless, experimental studies on the impurity states in EIs have been long absent.

## Results
Recently, Ta$_2$Pd$_3$Te$_5$ has been shown with the susceptibility to EI formation with macroscopic coherence[6,21-23], rendering it an appealing system for studying the impurity pair-breaking effects. Here, we present detailed microscopic evidence for the impurity-induced pair breaking in $Ta_2Pd_3Te_5$ with low-temperature scanning tunneling microscopy/spectroscopy (STM/S) at a base temperature of 1.6 K (see Supplementary Sec. I for experimental details). $Ta_2Pd_3Te_5$ is a van der Waals material with an orthorhombic lattice structure [space group: *Pnma*; Fig. 1(a)][24]. Within each of the two Te-Ta/Pd-Te trilayers in a unit cell, atoms are arranged in a quasi-one-dimensional (1D) manner, with Ta-Te or Pd-Te chains orientated along the *b* axis [Fig. 1(b)]. Figure 1(c) (upper panel) presents a typical topography of the cleaved (100) surface. The image displays periodic 1D ridges along *b* direction in the top Te layer. By decorating the STM tip with a $Ta_2Pd_3Te_5$ cluster, we can obtain the atomic-resolution pattern of five parallel Te chains (denoted as Te1–Te5) repeated along *c* axis [lower panel of Fig. 1(c)]. Fourier transformation of the surface Te lattice yields the lattice constants of $b_0$=3.70 Å and $c_0$=18.6 Å, in agreement with previous x-ray diffraction studies[25].

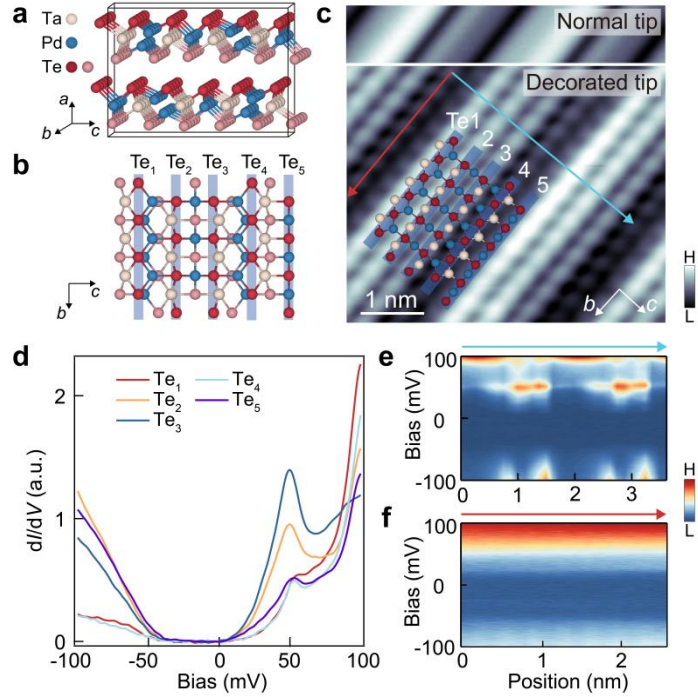

**Fig. 1 | Topography and tunneling spectra of $Ta_2Pd_3Te_5$.** (a,b) 3D and top views of the atomic structures of $Ta_2Pd_3Te_5$. Te1–Te5 chains (stripes) are marked. (c) Upper: topography of (100) surface. Lower: atomic-resolution STM image overlaid with the atomic structure. (d) Typical spectra collected on Te1–Te5 chains, respectively. (e,f) Spatially resolved spectra taken along the arrow-sketched trajectories in (c).

Tunneling-spectroscopy experiments are conducted to probe the local electronic structure. The d$I$/d$V$ tunneling spectrum, proportional to the local density of states (LDoS), exhibits a global insulating gap [Fig. 1(d)]. Previous STS and photoemission studies have shown that this gap is developed upon the excitonic-type hybridization of e–h bands[6,21,22]. Notably, the excitonic gap edge appears as pronounced coherence-like peak only for the electron-like (sample bias $V_b$>0) states at ~50 mV. By adopting the onset negative bias (~−35 mV) of non-zero d$I$/d$V$ as the hole-like gap edge, we obtain an under-estimated gap Δ of 85 meV, smaller than that measured by photoemission spectroscopy (~100 meV at 20 K)[6]. With a close examination of the spectra collected on Te1–Te5, we observe that despite the comparable gap-edge energies, their lineshapes vary due to the difference in local chemical environment. Such spectral variation is more clearly visualized in the spatially resolved spectra, where periodic spectral modulations can be seen normal to the Te chains [Fig. 1(e)], but absent along the Te chains [Fig. 1(f)], well corroborating the quasi-1D structure of $Ta_2Pd_3Te_5$.

Having identified the excitonic gap, we study the impurity scatterings for the possible excitonic pair breaking. Typical defects in $Ta_2Pd_3Te_5$ appear as elongated protrusions [Figs. 2(a) and S1]. Statistical analysis of the atomic-resolution imaging [for example, Fig. 2(b)], obtained via a tip decorated with $Ta_2Pd_3Te_5$ cluster, indicates that the defects are surface Te vacancies (assigned as $V_{Te1}$–$V_{Te4}$) mainly residing on Te1–Te4 chains (Fig. S2). Compared to the defect-free region, the spectra for all types of defects exhibit a prominent pair of resonance peaks near both sides of the gap edges [Fig. 2(c)]. To conclude the most representative spectral feature of $V_{Te1}$–$V_{Te4}$, we ordered and plotted with false color the typical spectra of all measured vacancies in Fig. S3, according to the energy difference $2\delta$ of electron ($E_e$, empty) and hole components ($E_h$, occupied) of the subgap states. The resulting $2\delta$-ordered spectra turn out to be basically grouped depending on the vacancy catalogs. Quantitatively, different types of vacancies show statistically distinct ranges of $2\delta$ [Fig. 2(d)], which present an increasing magnitude of the most probable $2\delta$ following an order of $V_{Te2} \rightarrow V_{Te4} \rightarrow V_{Te1} \rightarrow V_{Te3}$. Analogous to the impurity states in superconductors[26,27], the energy of excitonic pair-breaking states is supposed to be similarly dependent on the coupling between impurity and EI. Thus, the statistical separation in $2\delta$ for $V_{Te1}$–$V_{Te4}$ directly suggests that, the associated coupling strength between vacancy and Ta$_2$Pd$_3$Te$_5$ shows a tendency of dependence on the defect configuration. Such coupling-tunable energies of the impurity-induced subgap peaks highlight their origin from the e–h pair-breaking states[28-30].

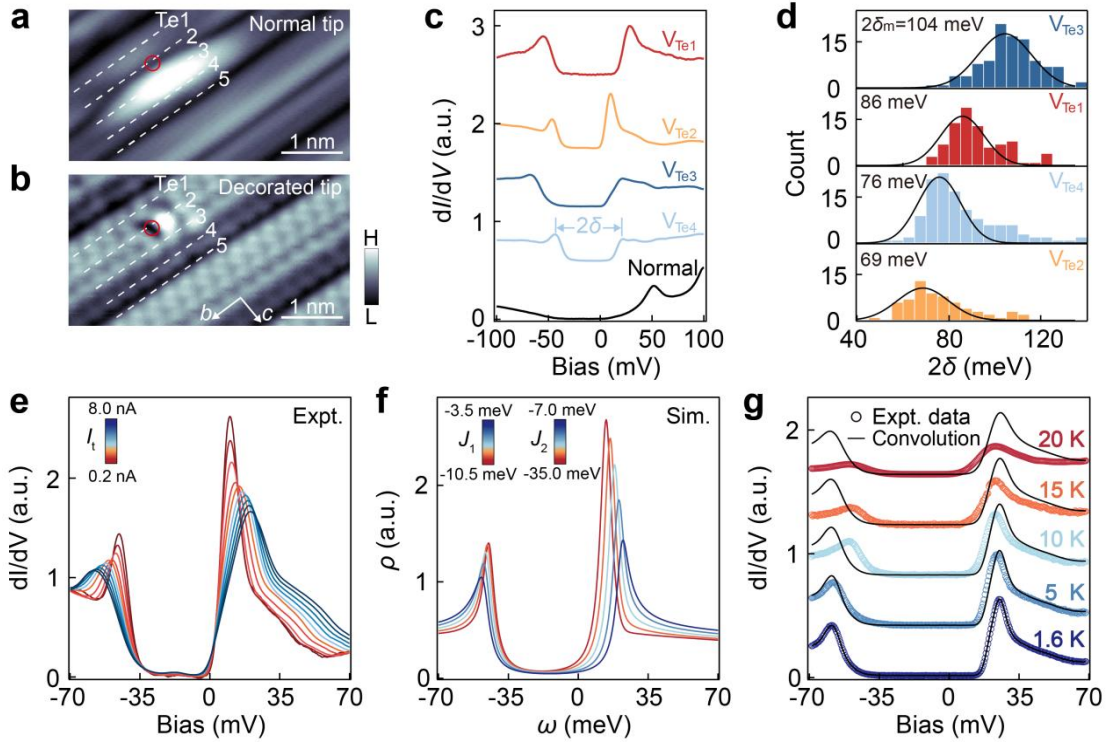

**Fig. 2 | Impurity-induced subgap states.** (a,b) Topographic images of a typical defect ($V_{Te2}$; red circle) without and with atomic resolution. (c) Spectra collected on defects $V_{Te1}$–$V_{Te4}$ (vertically offset). The 'normal' spectrum for defect-free region is plotted for comparison. (d) Histograms of $2\delta$ for $V_{Te1}$–$V_{Te4}$. The most probable $2\delta$, i.e., $2\delta_m$, for each type of defect obtained through Gaussian fitting (solid curve) is shown. (e) Normalized impurity spectra for $V_{Te2}$ measured with approaching the tip via increasing $I_t$ at fixed $V_b$=0.1 V (see Fig. S5 for reproducible data of $V_{Te1}$, $V_{Te3}$ and $V_{Te4}$). (f) Simulated excitonic impurity DoS spectra as decreasing $|J_1|$ and $|J_2|$ (zero-temperature excitonic gap $\Delta_0$=35 meV, chemical potential $\mu$=12.25 meV). (g) Measured excitonic impurity spectra on $V_{Te2}$ at different temperatures (vertically offset). The convoluted spectra at different temperatures are also plotted, which are obtained based on thermal convolution of the measured spectra at 1.6 K with Fermi function.

To study the pair-breaking properties more systematically, we explore the response of excitonic impurity states to external perturbations. As tip approaches the sample by increasing the tunneling current $I_t$, the subgap $E_{e,h}$ states at the defect (e.g., $V_{Te2}$) gradually shift towards the gap edge with broadened spectral peaks and suppressed intensity asymmetry [Figs. 2(e) and S4]. To explain the observation, we calculate the excitonic pair-breaking DoS using a two-band model of a singlet EI with a non-magnetic impurity (see Supplementary Sec. II for calculation details). In the model, electron scatterings that occur within the same band (intra-band scattering $J_1$) or in different bands (inter-band scattering $J_2$) are both considered. The sign of $J_1$ determines the asymmetry of subgap $E_{e,h}$ intensities: for $J_1<0$, the $E_e$ state exhibits stronger spectral weight than $E_h$ state, while the case is reversed for $J_1>0$ (Fig. S6). The strength of $J_2$ sets the energies $E_{e,h}$ of subgap states, with larger $J_2$ corresponding to smaller magnitude of $|E_{e,h}|$. By adopting attractive scattering potentials ($J_{1,2}<0$) with weaker intra-band component ($|J_1|<|J_2|$), the theory qualitatively reproduces the e–h imbalanced excitonic subgap states [Fig. 2(f) and Fig. S7], along with their spatial distributions (Fig. S8). Upon decreasing $|J_{1,2}|$, the calculated subgap $E_{e,h}$ states shift towards the gap edges, accompanied by peak broadening and weakened intensity imbalance [Fig. 2(f)], all in agreement with our measurements. Notably, the decrease in $|J_{1,2}|$ as tip approaching is unexpected, as the defect-induced potential should remain intact, while the tip-induced one should get enhanced. These calculations reveal the cooperative pair-breaking potentials with sign-opposite contributions from both of them, with the tip potential being relatively weaker.

Further check of the impurity-generated excitonic pair breaking is obtained by performing the temperature-dependent experiments. From the temperature evolution of the impurity spectra [Fig. 2(g)], we can see that the excitonic impurity subgap states shift toward the gap center and get suppressed in spectral intensity. Comparison with the spectra convoluted with Fermi function that incorporate pure thermal effect[31] suggests that the measured impurity spectra decay faster. This implies the existence of additional energy broadening mechanism, such as local suppression of the excitonic gap by tip potential and/or the quantum pairing 'poison'[32,33]. The accelerated decay, occurring alongside the thermal suppression of excitonic gap, agrees with the phenomenological behavior of impurity pair-breaking states[19]. Together with the theoretical reproduction of these subgap states under variable tip perturbations, the results consistently support the scenario of pair breaking for local excitons.

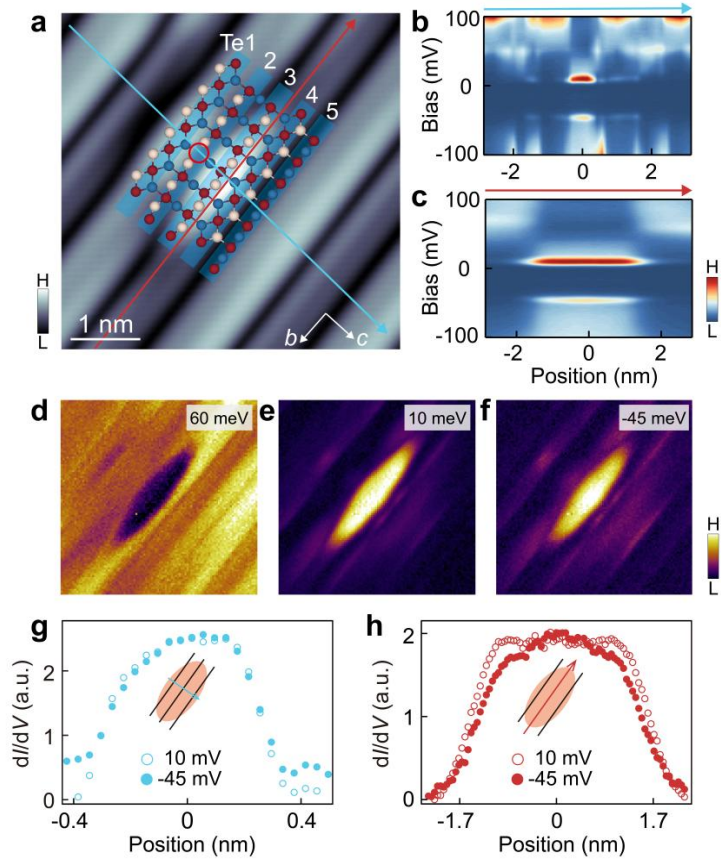

**Fig. 3 | Spatial distribution of the subgap states.** (a) Topographic image of defect $V_{Te2}$. (b,c) Spatially resolved spectra recorded across $V_{Te2}$ along the arrow-indicated trajectories in (a). (d–f) d$I$/d$V$ mappings for the $V_{Te2}$ defect in (a) taken at the biases of continuum (60 meV), $E_e$ (10 meV), and $E_h$ states (−45 meV). (g,h) d$I$/d$V$ linecuts crossing $V_{Te2}$ along $c$ (g) and $b$ directions (h) as depicted in the insets.

The next question naturally arises regarding how the pair breaking-related subgap states are distributed in space (Figs. 3 and S9). By measuring the spatially resolved spectra across the defect [Figs. 3(a)], we find that the subgap states show different spatial extensions along $c$ and $b$ axes [Figs. 3(b) vs. 3(c)], as dictated by the quasi-1D structure of Ta$_2$Pd$_3$Te$_5$. More directly, we collect d$I$/d$V$($r$, $E$) mappings at the biases of these subgap states for the same defect [Figs. 3(e) and 3(f)]. An elongated protrusion is explicitly resolved that closely follows the defect morphology. Notably, the topographic and spectral protrusions deviate from the vacancy site [circle in Fig. 3(a)]. This actually emerges from the

quasi-1D distortion of the predicated impurity states that are otherwise localized quasi-circularly near the vacancy outer edge[34]. In contrast, d$I$/d$V$ mapping for the electronic continuum (at 60 meV) outside the gap displays suppressed DoS near the defect site [Fig. 3(d)]. Such spectroscopic reversal, for states within and outside the gap, resembles that in pair-breaking superconductors[35]. It signifies the spectral weight transfer across the gap, suggesting the unique role of the excitonic gap in accommodating the gap-filling excitations. To describe the spatial distribution of subgap $E_{e,h}$ states, d$I$/d$V$($r$, $E_{e,h}$) are extracted crossing the defect along the $b$ and $c$ axes [Figs. 3(g) and 3(h)]. Quantitatively, along $c$ and $b$ directions, the topographic protrusion shows strikingly different widths (0.6 nm vs. 2.5 nm). However, in either direction, the d$I$/d$V$ linecuts for subgap $E_e$ and $E_h$ states collapse onto the same curve. This implies that $E_e$ and $E_h$ states are not independent, but inherently 'locked' as expected for pair-breaking subgap states in an EI[19,29,30].

We further probe coupled impurities to examine the electronic interactions between their pair-breaking states. Spectra of a $V_{Te}$ dimer of 1.3 nm apart reveal splitting of the subgap states [Figs. S10(a) and S10(b)]. This feature is intrinsic to the dimer with internal interactions, as it is absent in a similar dimer structure yet with a sufficiently large inter-$V_{Te}$ separation (~2 nm, Fig. S11). Differential conductance maps at the corresponding energies of these splitted sub-gap states exhibit the spatial distributions that resemble the hybridized bonding and antibonding orbitals [Figs. S10(c)–S10(f)][36]. The results for the interacting dimer demonstrate the electronic coupling of quantum pair-breaking states.

Finally, we study the influence of electronic inhomogeneity. On the Ta$_2$Pd$_3$Te$_5$ surface, charge inhomogeneous regions manifest in spectra as distinctly separated energies of a characteristic normal-state peak [black arrows, Fig. 4(b)], defined as type-I and type-II regions, respectively. Since the excitonic gap center lies below the Fermi level [Fig. 4(b)], this implies intrinsic e–h imbalance in Ta$_2$Pd$_3$Te$_5$ with excess of electrons. Compared with type-I region, the type-II one with a smaller characteristic-state energy is more electron-doped, thus exhibiting stronger e–h imbalance. Defects in type-I regions favor subgap $E_{e,h}$ states near gap edges. In contrast, defects in the more e–h imbalanced type-II regions host subgap states that are much closer to the Fermi level [Fig. 4(c)], referred to as lower-energy subgap $E_{e,h}$ states. In space, the spectra showing higher-energy subgap states frequently appear near the dark regions and its surroundings in topography [Fig. 4(a)]. However, those exhibiting lower-energy subgap states tends to emerge in the bright regions. These observations highlight the crucial role of spatial charge inhomogeneity in shaping the impurity states.

During spectra measurement, we find that increasing the tunneling current $I_t$ beyond a sufficiently large level (≳25 nA) can 'trigger' and amplify the lower-energy subgap state [Fig. 4(e)]. Further raising $I_t$ then drives both higher- and lower-energy states toward the electronic continuum outside the gap. Furthermore, the lower-energy subgap states show a far lower surviving temperature than the higher-energy ones (Fig. S12). These behaviors identify the lower-energy subgap state as another type of pair-breaking effect. Theoretically, in regions where e–h imbalance is stronger than their surroundings, the excitonic gap is more susceptible to pair breaking, and thus would be locally suppressed. By considering such spatial gap inhomogeneity in our mean-field theory, the lower-energy subgap state can be generated, as a result of the reduced excitonic order parameter at the impurity site [Figs. 4(d) and S13], and coexist with the higher-energy subgap states [Fig. 4(f)]. Their coexistence demonstrates the co-tunneling through different paths associated with locally suppressed and unperturbed excitonic order parameters.

While the similarities of EIs to superconductors have motivated us to conclude the pair breaking, their difference in several important aspects should be noted in future explorations. i) Spectroscopic asymmetry. The excitonic gapped spectrum, together with the impurity-induced subgap states, exhibits the absence of e–h symmetry that is instead present in superconductors, as the excitonic pairing does not require particle–hole symmetry. ii) DoS distribution. Unlike the situation of superconductors[37], the spatial distribution of impurity subgap states in EIs shows no obvious structure of a specific orbital, and thus should be contributed by the mixture of multiple orbitals. The associated preservation of orbital degeneracy in EIs is likely because the crystal field splitting involved, typically ~100 times smaller than the energy of excitonic gap, is negligible. iii) Impurity sensitivity. In superconductors, the Cooper depairing is impurity-selective, that is, only magnetic impurities exhibit pair breaking in spin-singlet pairing as required by the Anderson's theorem[19]. However, in EIs, both of magnetic and nonmagnetic impurities show the pair-breaking effects, regardless of whether the pairing is spin-singlet or -triplet. This means that to determine the concrete pairing symmetry in EIs, further studies beyond the conventional impurity-based probes are needed, for example, via tracking the quantitative evolution of subgap states with impurity concentration[30].

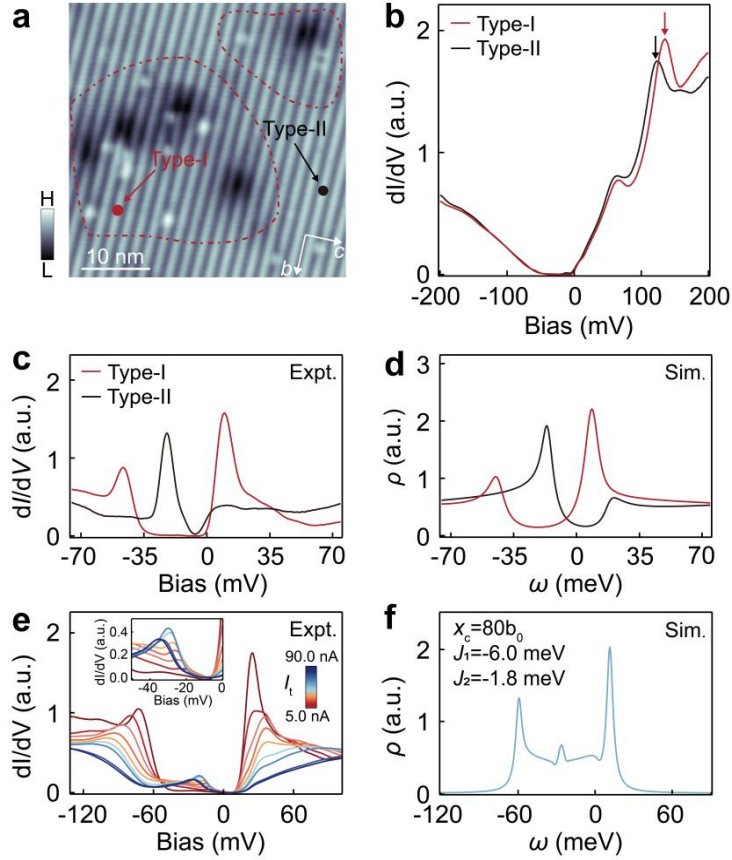

**Fig. 4 | Interplay between subgap states arising from different excitonic order parameters.** (a) Topographic image showing the electronic inhomogeneity. The dashed curve outlines regions exhibiting relatively stronger e–h balance. (b) Large-energy spectra taken at different defect-free regions as exemplified by dots in (a). (c) Typical spectra acquired on $V_{Te2}$ in type-I and type-II regions in (a), showing higher- and lower-energy subgap $E_{e,h}$ states, respectively. (d) Simulated impurity DoS with higher- ($\Delta_0$=45 meV, $J_1$=−15 meV, $J_2$=−40 meV) and lower-energy subgap states ($\Delta_0$=18 meV, $J_1$=20 meV, $J_2$=−10 meV). (e) Normalized impurity spectra with coexisting higher- and lower-energy subgap $E_{e,h}$ states under increasing $I_t$ at fixed $V_b$=0.1 V. Inset shows a magnified view of the low-energy spectra. Note that the $E_e$ component of lower-energy subgap states otherwise expected at $V_b$>0 has merged into the $E_e$ branch of higher-energy subgap states. (f) Calculated impurity DoS with a spatially inhomogeneous excitonic gap $\Delta(x) = \Delta_0 \tanh\left(\frac{x}{x_c}\right)$ ($\Delta_0$=30 meV).

**Conclusion**

In summary, we demonstrate via tunneling spectroscopy the atomic characterization of pair breaking in $Ta_2Pd_3Te_5$ as a local probe of the depairing excitonic condensate. The results in turn confirm that the probed insulating gap at impurity-free region arises from the excitonic pairing. Our findings show the prospect of developing an impurity scattering-based formulism[19] for investigating the microscopic establishment of global excitonic coherence. The shared nature both as bosonic condensate for EIs and superconductors highlights a wide area beyond the impurity scatterings to explore, particularly the excitonic versions of gap function, vortex in superfluid, and Josephson effect. These researches of EI complementary to superconductors would map out a full picture of the fermion-pair condensate, which is helpful for a thorough understanding of various bosonic correlated phenomena, such as excitonic Mott insulator[38], excitonic spin superfluid[39], and the competing phase diagram of high-$T_c$ superconductivity[40].

*Note added*: During the review of our manuscript, we become aware of a similar study[41].

**Data availability**

Source data are provided with this paper. Any additional material is available from the corresponding authors upon reasonable request.

**Acknowledgments**

This work is funded by the National Key Research and Development Program of China (Grants No. 2022YFA1402400, 2022YFA1403601, 2024YFA1410500, and 2024YFA140840), the National Science Foundation of China (Grants No. 92265201, 92477137, 12174131, 12404210, 12322402, and 12274206), the Innovation Program for Quantum Science and Technology (Grant No. 2021ZD0302800), the Natural Science Foundation of Jiangsu Province (Grants No. BK20233001), the Xiaomi foundation, and the Synergetic Extreme Condition User Facility (SECUF, https://cstr.cn/31123.02.SECUF).

**Author contributions**

Y.-S.F and C.L. supervised the project. X.D. and Y.S. prepared the single crystal. L.Y., H.Z., Y.Z. and C.Z. carried out the STM experiments under the supervision of Y.-S.F., C.L. and W.Z. D.W. performed the calculations under the supervision of R.W. C.L., Y.-S.F., and R.W. prepared the manuscript, with comments from all authors.

**Competing interests**

The authors declare no competing interests.